\documentclass[11pt,a4paper]{article}
\usepackage[T1]{fontenc}
\usepackage[utf8]{inputenc}
\usepackage{amssymb,amsmath}
\usepackage[dvips]{graphicx}
\usepackage{mathrsfs}
\usepackage{setspace}
\usepackage{indentfirst}
\def\la{\lambda}
\def\ka{\kappa}
\def\om{\omega}

\def\si{\sigma}
\def\ro{\varrho}
\def\de{\delta}

\def\ep{\varepsilon}
\def\pa#1#2{\partial{#1}/\partial{#2}}
\def\Pa#1#2{\frac{\partial{#1}}{\partial{#2}}}

\def\Ppa#1#2#3{{\frac{\partial^2{#1}}{\partial{#2}\partial{#3}}}}

\def\po{{\frac{1}{2}}}

\def\yp{\upsilon}

\begin{document}

\bibliographystyle{unsrt}

\title{\bf Plane waves in a relativistic\\homogeneous and isotropic
elastic continuum}
\author{Vratko Pol\'{a}k\footnote{e-mail address:
polak@fmph.uniba.sk}\ \ and Vladim\'\i r Balek\footnote{e-mail
address: balek@fmph.uniba.sk}
\\
{\it Department of Theoretical Physics,
Comenius University, Mlynsk\'{a} dolina}
\\
{\it 842 48 Bratislava, Slovakia}}

\maketitle
\maketitle\abstract

{Propagation of gravitational and acoustic plane waves in a flat
universe filled with a general relativistic, homogeneous and
isotropic, spatially flat continuum is studied. The continuum is
described by analogues of nonrelativi\-stic characteristics,
namely energy per particle, pressure and Lame coefficients, and
considered in the comoving proper--time gauge. For all modes with
the given wave covector, differential equations governing the time
dependence of the amplitudes are derived. In particular,
longitudinal acoustic waves are described, in analogy with the
nonrelativistic theory, by two coupled first--order equations. As
an example, plane waves in a stiff ultrarigid continuum are
considered.}

\section{Introduction}\label{i}

Elasticity, as a theory of matter in form of an (ideal,
dissipationless) elastic continuum, has many applications in
everyday life; but in areas where relativistic effects are
important, applications are much more sparse. Nevertheless, the
presence of a solid crust in neutron stars has motivated the
development of relasticity (short name for relativistic
elasticity, from \cite{KS1}). Among early works on the topic,
\cite{CQ} appears to be the most successful.

Since then, fundamental concepts of relasticity are known, but the
terminology, notation and formalism are far from being stabilized.
Different authors start from different notions of nonrelativistic
elasticity, and use different coordinates, depending on what they
find most natural and/or appropriate for the problem they study.
This paper is no exception. We deal with a particular problem,
thus we consider particular notions to be basic, develop
particular formalism and perform computations in particular
coordinates.

The motivation of this paper is that, as long as elastic continuum
can be included into the framework of general relativity, we
should be able to describe how nonrelativistic elastic phenomena
and Einstein's gravitational phenomena affect each other.

As a nonrelativistic effect to start with, we have chosen the
propagation of acoustic waves; specifically, its simplest case,
the propagation of plane waves in a homogeneous and isotropic
continuum. For details, see \cite{Brdicka}.

In general relativity, the effects described by the theory also
include plane waves in a homogeneous and isotropic background,
namely gravitational waves in a flat expanding universe.

The first goal of this paper is to derive equations of propagation
of both acoustic and gravitational (weak) waves in a homogeneous
and isotropic, spatially flat, but otherwise completely general
continuum. Since less specific equations for general weak waves
were formulated, and the correctness of their nonrelativistic
limit was proven in \cite{CQW}, we do not address the problem of
nonrelativistic limit here.

The second goal of this paper is to show how to use the theory of
relasticity when solving a particular problem, if simplicity and
clarity are considered more important than elegance and
universality. This goal includes introducing such formalism, and
using such coordinates, that the result will be obtained in the
easiest possible way, even if that formalism and those coordinates
will not prove useful in other relasticity problems.

This paper is motivated neither by astrophysics nor by cosmology,
although its results can be applied within the latter, when
considering a solid dark matter component as in \cite{SDM}.
Therefore, when describing our results, we remind the reader also
of the terminology used in the theory of cosmological
perturbations, see for example \cite{LiLy}. Our notations in
relasticity are chosen to be compatible with \cite{Beig}, and our
notations in differential geometry to be compatible with
\cite{Fecko}.

The paper is organized as follows. In section \ref{r} we introduce
basic concepts of relasticity. In section \ref{pw} we develop a
theory of plane waves as linear perturbations imposed on the
background solution. We derive equations for the background as
well as for the perturbations, identify modes that correspond to
gravitational and acoustic (transversal and longitudinal) waves,
and propose a gauge invariant description of longitudinal acoustic
waves. For better understanding of our formalism, in section
\ref{e} we write down equations for plane waves in a particular
continuum, namely the stiff ultrarigid continuum introduced in
\cite{KS3}. In section \ref{z} we discuss the results.

\section{Relasticity}\label{r}

Relasticity is a theory that describes elastic continua within
Einstein's theory of gravity. In the simplest case there is just
one continuum, and it extends over all spacetime. The spacetime
is, of course, a 4--dimensional manifold, equipped with a metric
$g$ that is regarded as a dynamical field satisfying the Einstein
equations.

Properties of the continuum are defined on the material manifold,
that is, on a 3--dimensional manifold whose points represent
particles of which the continuum consists. The way the continuum
is distributed in spacetime is described by the mapping $f$ that
maps spacetime on the material manifold, telling us which particle
is in the particular worldpoint. It is called material mapping and
considered a dynamical field on the spacetime, too. Contravariant
metric in any worldpoint can be push--forwarded by the material
mapping, providing us with a symmetric, double contravariant
tensor on the material manifold. This tensor can be interpreted as
the strain tensor.

The only characteristic of continuum that influences its dynamics
is the state function $\ro$. It is a scalar function with two
arguments, the point of the material manifold and the strain
tensor at that point. By definition, it is the energy density in
the rest frame of the given particle, in the continuum whose
deformation is described by the given strain tensor.

The dynamics is determined by the action $S = \displaystyle \int
(- \rho + R) \omega_g$, where $R$ is the scalar curvature, $\rho$
is the energy density given by the state function and $\om_g$ is
the 4--form of volume defined by the metric $g$ (for details, see
\cite{Fecko}). We use a system of units in which the factor in
front of $R$ involving gravitational constant is absent. In these
units, the Einstein equations assume the form $2G_{\mu\nu} =
T_{\mu\nu}$.

The principle of extremal action provides us with the equations of
motion. Variation of the action with respect to the metric leads
to the Einstein equations with an explicit formula for the
energy--momentum tensor (to be given in the next section).
Variation of the action with respect to the material mapping
yields equations of motion of the continuum (also written in the
next section). As a matter of fact, Einstein equations together
with the Bianchi identity usually carry enough information to
describe the dynamics completely; nevertheless, it is useful to
write down the equations governing the dynamics of the source
explicitly.

The action, and therefore also equations of motion, is covariant
with respect to the change of coordinates in the spacetime, and
also in the material manifold. This is quite a large gauge
freedom. This freedom, however, can be used for fixing
coordinate--dependent expressions in such a way that they become
simpler (although less general). In this paper, we fix the gauge
almost completely in a way we find most suitable for our problem.

\section{Plane waves}\label{pw}
We will investigate the propagation of weak waves, which can be
regarded as linear perturbations to a certain background solution.

\subsection{Description of the continuum; the background solution}
\label{pw;bs}
For simplicity, we choose the elastic continuum to be homogeneous,
isotropic and flat, which means that there exists a set of
coordinates $X^A$ in the material manifold (where the material
index $A$, as well as any other upper case Latin index, runs from
$1$ to $3$) such that the state function is invariant with respect
to Euclidean translations and rotations of $X^A$.

The strain tensor has coordinate components (with respect to
$X^A$) which can be treated as coordinates on the space of double
contravariant tensors on the material manifold. These components
will be denoted by $H^{AB}$. The true strain tensors are always
symmetric, but for the purposes of differentiation we can extend
the state function also to nonsymmetric tensors. This gives
correct results if we let $\ro$ depend only on the symmetric part
of $H^{AB}$, or if we symmetrize, when differentiating with
respect to $H^{AB}$ more than once, the derivatives with respect
to all $H^{AB}$'s but the last (and of course, insert symmetric
values into the resulting expressions). We will adopt the latter
procedure. In general, $\ro$ can be differentiated also with
respect to $X^A$. However, the invariance with respect to
translations implies $\pa{\ro}{X^A} = 0$, therefore only the
derivatives of $\ro$ with respect to (the nine components of)
$H^{AB}$ enter the equations.

Having fixed the coordinates $X^A$ on the material manifold, we
can express the material mapping $f$ through the coordinate
component functions $f^A$. Together with arbitrary coordinates
$x^\mu$ in the spacetime (where the spacetime index $\mu$, as well
as any other Greek index, runs from 0 to 3; we will also use lower
case Latin letters to denote space indices running from 1 to 3),
we now have everything ready for the variation of the action with
respect to $f^A$. This variation leads to the equations of motion
of a homogeneous isotropic flat continuum
\begin{equation}
\left(\Pa{\ro}{H^{AB}}g^{\mu\nu} +2\Ppa{\ro}{H^{AC}}{H^{BD}}
g^{\mu\la}g^{\nu\ka}f^C_{,\la}f^D_{,\ka}\right)f^B_{,\nu;\mu} =0.
\label{eq:eqofmotion}
\end{equation}
By varying the material part of the action with respect to
$g_{\mu\nu}$, we obtain a formula for the energy--momentum tensor,
\begin{equation}
T_{\mu\nu} = 2\Pa{\ro}{H^{AB}}f^A_{,\mu}f^B_{,\nu} -\ro
g_{\mu\nu}. \label{tensor}
\end{equation}

Having completed a general description of the continuum under
consideration, we can proceed to further simplifications. The
gauge freedom in spacetime coordinates can be reduced by requiring
that $f^A(x) = x^A$ ($x^A$ are comoving coordinates), after which
there remains only one free coordinate $x^0$. Further restriction
will be $g_{00} = -1$, which implies that $x^0$ equals the proper
time $t$ of the particles of the continuum. The only freedom left
consists in the choice of the hypersurface $x^0 = 0$.

Now we can specify the background. We assume that it is, just like
the continuum, homogeneous, isotropic and flat, that is, equipped
with a flat Robertson--Walker metric $g = -dt\otimes{dt}
+a^2(dx^1\otimes{dx^1} +dx^2\otimes{dx^2} +dx^3\otimes{dx^3})$,
where the scale parameter $a$ depends only on $t$. The symmetry
allowed us to choose the zero time hypersurface to be orthogonal
to the worldlines of all particles. Instead of the scale
parameter, it turns out to be more convenient to use its logarithm
$z = \ln{a}$. Then we have $g_{00} = -1$, $g_{0i} = 0$, $g_{ij} =
e^{2z} \de_{ij}$, and we can compute the Einstein tensor to obtain
$2G_{00} = 6\dot{z}^2$, $2G_{0i} = 0$, $2G_{ij} =
e^{2z}(-4\ddot{z}-6\dot{z}^2) \de_{ij}$, where the overdot means
differentiation with respect to $t$. The energy--momentum tensor
simplifies to $T_{00} = \ro$, $T_{0i} = 0$, $T_{ij} =
2\pa{\ro}{H^{ij}} -e^{2z}\ro\de_{ij}$. Equations of motion of the
continuum hold identically.

The strain tensor of the background solution is $H^{AB} =
e^{-2z}\delta^{AB}$, and the invariance of $\ro$ with respect to
rotations ensures that 
the derivatives of $\ro$ with respect to $H^{AB}$, evaluated for
the background solution, are proportional to $\de_{AB}$ (the first
derivative) or equal to linear combinations of the products of
$\de_{AB}$ (the higher derivatives). Here, $\de$ is in fact the
Euclidean metric of the material manifold, but in our coordinates
we can use notation reminding us that $\de_{AB}$ and $\de^{AB}$
are unit matrices. This enables us to employ an extended version
of summation convention, for example $g^{AA} = g^{AB}\de_{AB}$ and
$g_{AA} = g_{AB}\de^{AB}$. Since we have, in fact, identified the
spatial coordinates with the material coordinates, we can
pull--back $\de$ to obtain the unit matrices $\de_{ij}$ and
$\de^{ij}$. The first matrix was already used in the formulas for
$2G_{ij}$ and $T_{ij}$.

As the next step, let us introduce material characteristics to
appear in the equations for the background, as well as in the
equations for perturbations. First, Einstein's cosmological
constant can be treated as the vacuum energy density $\rho_0$
contributing a constant term to $\ro$. After separating out this
term, we can write the remaining energy density as a product of
the particle density $n$ and the energy per particle $\ep$. In
this way we obtain the general formula $\ro= \rho_0 + n\ep$, which
can be written for the background metric as $\ro = \rho_0 +
e^{-3z}\ep$. Note that the separation of the vacuum energy density
is in a sense artificial. We could choose any other value of
$\rho_0$ and change $\ep$ in such a way that $\ro$ will be the
same. Nevertheless, the introduction of $\rho_0$ is useful since
it provides us with a simpler formula for the quantity $\ep$ than
we would obtain if we put simply $\ro= n\ep$.

In the nonrelativistic theory, when a homogeneous and isotropic
continuum is deformed in a homogeneous and isotropic way, its
stress tensor is given by one scalar quantity -- the pressure
$\sigma$; and when it is relaxed, its elastic properties are
determined by two scalar quantities -- Lame coefficients $\mu$ and
$\la$ (see, for example, \cite{Brdicka}). Analogically, for a
general relativistic homogeneous and isotropic continuum we define
the scalar $\si$ by the relation
$$2\Pa{\ep}{H^{AB}} = e^{2z}\si\de_{AB},$$
and the scalars $\mu$ and $\la$ by the relation
$$2\left(\Ppa{\ep}{H^{AB}}{H^{CD}} +\Ppa{\ep}{H^{AB}}{H^{DC}}\right) =
e^{4z}\left[\la\de_{AB}\de_{CD} +\mu(\de_{AC}\de_{BD}
+\de_{AD}\de_{BC})\right].$$ With all this new scalars, the
derivatives of $\ro$ evaluated for the background metric can be
written as
\begin{equation}
2\Pa{\ro}{H^{AB}} = e^{-z}(\ep +\si)\de_{AB} \label{dro}
\end{equation}
and
\begin{equation}
2\left(\Ppa{\ro}{H^{AB}}{H^{CD}}+\Ppa{\ro}{H^{AB}}{H^{DC}} \right)
= e^z[(\la +2\si +\ep)\de_{AB}\de_{CD} +(\mu
-\ep)(\de_{AC}\de_{BD} +\de_{AD}\de_{BC})]. \label{ddro}
\end{equation}
The scalars $\ep$, $\si$, $\mu$ and $\la$ as functions of $z$ are
not independent, because the derivative with respect to $z$
corresponds to the derivative with respect to $H^{AB}$. The
relations between these scalars are ${d\ep}/{dz} = -3\si$ and
${d\si}/{dz} = -(2\si +3\la +2\mu)$.

Now we are ready to write down the equations for the background.
$T_{00}$ becomes $\rho_0 + e^{-3z}\ep$, which gives $6\dot{z}^2 =
\rho_0 +e^{-3z}\ep$, and $T_{ij}$ becomes $e^{-z}\si\de_{ij}
-e^{2z}\rho_0\de_{ij}$, which gives, when combined with the
equation for $\dot{z}^2$, $-4\ddot{z} = e^{-3z}(\ep + \si)$. If we
take into account the identity for ${d\ep}/{dz}$, we can see that
the latter equation is nothing but the time derivative of the
former equation. (At least, this holds if $\dot{z} \neq 0$, which
we will assume in what follows in order to avoid singularities in
our expressions. By the same reason, we will assume $\ep +\si \neq
0$.) Thus, the background is a flat Robertson--Walker metric
satisfying $6\dot{z}^2 = \rho_0 +e^{-3z}\ep$ and ${d\ep}/{dz} =
-3\si$; in other words, a flat universe evolving as if it
contained, in addition to the dark energy with the density
$\ro_0$, an ideal fluid with the density $e^{-3z}\ep$ and pressure
$e^{-3z}\si$.

\subsection{Linear perturbations}\label{pw;lp}
Having fixed the background solution we can start to inspect
linear perturbations. Denote perturbations of the fields $f$ and
$g$ by $\de{f}$ and $\de{g}$. Of course, not every perturbation is
permitted by the dynamics, but on the other hand, some
perturbations correspond merely to coordinate changes and are not
of physical interest.

First, we reduce the gauge degrees of freedom in a way analogical
to how we fixed the coordinates in the background solution. We
require that for the perturbed solutions the spacetime coordinates
are comoving, too, which implies $\de{f}^A = 0$. We want also the
$x^0$ coordinate to measure the proper time of the particles ($x^0
= t$), which implies $\de{g}_{00} = 0$. This leaves us with the
freedom of choosing the zero time hypersurface. But, unlike the
background, the perturbed metric is generally not symmetric enough
to allow us to pick a certain nice hypersurface (like the
hypersurface orthogonal to the worldlines of all particles).
Therefore we proceed further with this last ambiguity left.

Before writing down the linearized dynamical equations, we can
take advantage of the fact that the linearized equations are (of
course) linear. Therefore their solutions form a linear space
(they satisfy the superposition principle), and can be
complexified. The complex solutions have no direct physical
meaning, but they can simplify the upcoming expressions a bit.

Since the solutions form a linear space, it is sufficient to
investigate some basis of them to obtain the complete picture. A
nice ansatz for such basis follows from the translation symmetry
of the continuum. This symmetry ensures that there are solutions
of the form of (complex) plane waves, that is, depending on the
space coordinates only through the multiplication factor $\exp(i
k_A x^A)$, where $k$ is a constant real wave covector on the
material manifold. When dealing with just one plane wave, we can
always utilize the rotational invariance of the background
solution to rotate coordinates in such a way that $x^1$ becomes
parallel to $k$, which means that the coordinate components of $k$
become $(k, 0, 0)$ (the first $k$ is a covector, the second $k$ is
a real number).

If we find all plane wave solutions, we can use them to construct
any solution of the form of a (space--) tempered distribution.
This means that we actually do not have a basis, because we miss
the solutions that grow faster than polynomially with the space
distance, as is the case for surface waves. We duly restrict our
attention to the space--tempered distributions, referring to the
fact that because of causality such solutions are sufficient for
describing linearized dynamics in a bounded region of spacetime.

Plane waves provide a significant simplification of our problem
since the space derivatives acting on the perturbations become
just powers of $k$. In further computations, some typical
combinations of derivatives with respect to time or space
coordinates occur repeatedly, so it is convenient not to use
$\de{g}$ directly, but to replace it by more complicated
expressions in order to obtain simpler looking results. For a
plane wave with the wave covector $k$ already rotated into the
$x^1$ direction we define a tensorial--looking, time--dependent
quantity $h$ by putting
$$\de g_{0i}(x) = -i k h_{0i}(t)e^{ikx^1},$$
$$\de g_{ij}(x) = e^{2z(t)}h_{ij}(t)e^{ikx^1}.$$

Now, everything is ready for a straightforward but lengthy
variation of dynamical equations. We will write down just the
final expressions. Variation of the energy--momentum tensor gives
$$\de T_{00} = -\po e^{-3z}(\ep +\si)h_{kk},$$
$$\de T_{0i} = ik \rho_0 h_{0i} +ik e^{-3z}\ep h_{0i},$$
$$\de T_{ij} = -\po e^{-z}(\la +\si)\de_{ij}h_{kk}
-e^{-z}\mu h_{ij} -e^{2z}\rho_0 h_{ij};$$ variation of the
Einstein tensor gives
$$2\de G_{00} = 2\dot{z}\dot{h}_{kk} -4e^{-2z}k^2
\dot{z}h_{01} +e^{-2z}k^2(h_{kk} -h_{11}),$$
$$2\de G_{01} = ik\dot{h}_{11} -ik\dot{h}_{kk} +ik(4\ddot{z}
+6\dot{z}^2)h_{01},$$
$$2\de G_{0\alpha} = ik\dot{h}_{1\alpha} +ik(4\ddot{z} +6\dot{z}^2)h_{0\alpha}
-ike^{-2z}k^2 h_{0\alpha},$$
$$2\de G_{11} = e^{2z}[\ddot{h}_{11}
+3\dot{z}\dot{h}_{11} -(4\ddot{z} +6\dot{z}^2)h_{11}
-\ddot{h}_{kk} -3\dot{z}\dot{h}_{kk}],$$
$$2\de G_{1\alpha} = e^{2z}[\ddot{h}_{1\alpha}
+3\dot{z}\dot{h}_{1\alpha} -(4\ddot{z} +6\dot{z}^2)h_{1\alpha}]
-k^2 \dot{h}_{0\alpha} -k^2\dot{z}h_{0\alpha},$$
$$2\de G_{22} = e^{2z}[\ddot{h}_{22} +3\dot{z}\dot{h}_{22}
-(4\ddot{z} +6\dot{z}^2)h_{22} -\ddot{h}_{kk}
-3\dot{z}\dot{h}_{kk}] +2k^2\dot{h}_{01} +2k^2\dot{z}h_{01}
+k^2(h_{11} +h_{22} -h_{kk}),$$
$$2\de G_{33} = e^{2z}[\ddot{h}_{33} +3\dot{z}\dot{h}_{33}
-(4\ddot{z} +6\dot{z}^2)h_{33} -\ddot{h}_{kk}
-3\dot{z}\dot{h}_{kk}] +2k^2\dot{h}_{01} +2k^2\dot{z}h_{01}
+k^2(h_{11} +h_{33} -h_{kk}),$$
$$2\de G_{23} = e^{2z}[\ddot{h}_{23} +3\dot{z}\dot{h}_{23}
-(4\ddot{z} +6\dot{z}^2)h_{23}] +k^2h_{23};$$ and variation of the
equation of motion of the continuum gives
$$(\ep +\si)\dot{h}_{01} = (2\mu +3\la +5\si)\dot{z}h_{01}
-(\mu +\si)h_{11} -\po(\la +\si)h_{kk},$$
$$(\ep +\si)\dot{h}_{0\alpha} = (2\mu +3\la +5\si)\dot{z}h_{0\alpha}
-(\mu +\si)h_{1\alpha}.$$ The index $\alpha$, appearing in the
expressions for $\de G_{\mu \nu}$ as well as in the varied
equations of motion of the continuum, assumes values 2 and 3. The
expressions for $\de T_{\mu \nu}$ and $2\de G_{\mu \nu}$ are given
modulo factor $e^{ikx^1}$; this is, however, of no significance
since we aim to equate them anyway.

When writing down the varied Einstein equations we can use
formulas for the background solution to eliminate both $\rho_0$
and $\ddot{z}$ from our equations (and keep $z$, $\dot{z}$,
$\ep+\si$, $\la+\si$ and $\mu+\si$ only). Again, as in the
background solution, we find that some terms of the varied
Einstein equations are just time derivatives of other equations.
In addition to the varied equations of motion of the continuum
cited above, we obtain the following set of independent equations:
$$2\dot{z}\dot{h}_{kk} +e^{-2z}k^2(h_{kk} -h_{11} -4\dot{z}h_{01}) =
-\po e^{-3z}(\ep +\si)h_{kk},$$
$$\dot{h}_{11} -\dot{h}_{kk} = e^{-3z}(\ep +\si)h_{01},$$
$$\dot{h}_{1\alpha} -e^{-2z}k^2 h_{0\alpha} = e^{-3z}(\ep +\si)
h_{0\alpha},$$
$$(\ddot{h}_{22} -\ddot{h}_{33}) +3\dot{z}(\dot{h}_{22} -\dot{h}_{33})
+e^{-2z}k^2(h_{22} -h_{33}) = -e^{-3z}(\mu +\si)(h_{22}
-h_{33}),$$
$$\ddot{h}_{23} +3\dot{z}\dot{h}_{23} +e^{-2z}k^2h_{23} =
-e^{-3z}(\mu +\si)h_{23}.$$

\subsection{Modes}\label{pw;m}
The set of equations we have found is evidently not totally
coupled; it can be rather divided into several subsets of coupled
equations. Physically, this means that there are different modes
of wave propagation for the given wave covector. Since general
relativity without matter and classical elasticity are both limit
cases of relasticity, we expect that there will be two modes of
gravitational waves, two modes of transversal acoustic waves and
one mode of longitudal acoustic waves. Also, there should remain
one odd mode coming from the freedom of choosing the zero time
hypersurface.

First we have equations for $h_{22} -h_{33}$ and $h_{23}$, which
can be written in a compact form as
\begin{equation}
\ddot{h}^T_{ij}+3\dot{z}\dot{h}^T_{ij}+ \left[e^{-2z}k^2 +
e^{-3z}(\mu +\si)\right]h^T_{ij} = 0, \label{grav}
\end{equation}
where $h^T_{ij}$ is the tensor part of $h_{ij}$ (a symmetric 3
$\times$ 3 matrix with nonzero components $h^T_{22} = - h^T_{33} =
(h_{22} - h_{33})/2$ and $h^T_{23} = h_{23}$). These equations
clearly describe gravitational waves with polarizations $\oplus$
(the component $h^T_{22}$) and $\otimes$ (the component
$h^T_{23}$); or, speaking in cosmological terms, the tensor
perturbations (see \cite{LiLy}). The equations coincide with those
valid in general relativity for an universe filled with an ideal
fluid, see \cite{LaLi}, if $\mu +\si$ vanishes. To see where this
condition comes from, note that for an ideal fluid the energy per
particle $\ep$ depends only on the particle density $n$. If we
write $n = \displaystyle \sqrt{\mbox{det} H^{AB}}$, we can compute
the second derivatives of $\ep = \ep (n)$ with respect to $H^{AB}$
and compare them with the expression in subsection \ref{pw;bs}. In
this way we obtain $\mu = - n d\ep/dn = (1/3) d\ep/dz$. However,
the identity for $d\ep/dz$, to be found also in subsection
\ref{pw;bs}, implies that the right hand side equals $-\si$. Thus,
an ideal fluid can be defined as an elastic continuum with $\mu
+\si = 0$.

Next we have two pairs of equations containing $h_{0\alpha}$ and
$h_{1\alpha}$ only,
\begin{equation}
\dot{h}_{0\alpha} = \frac{2\mu +3\la +5\si}{\ep +\si}\dot{z}
h_{0\alpha}-\frac{\mu +\si}{\ep +\si}h_{1\alpha}, \label{trans1}
\end{equation}
\begin{equation}
\dot{h}_{1\alpha} =\left[e^{-2z}k^2 +e^{-3z}(\ep
+\si)\right]h_{0\alpha}. \label{trans2}
\end{equation}
Consider equations for $h_{02}$ and $h_{12}$. Since the
perturbation of $g_{12}$ represents the shear strain in the $x^2$
direction and the perturbation of $g_{02}$ represents the motion
of the particles of the continuum in the same direction (for an
observer for whom the time $t$ is locally synchronized), this mode
can be identified with the transversal acoustic wave that
oscillates in the $x^2$ direction. Analogically, equations for
$h_{03}$ and $h_{13}$ describe the propagation of a transversal
acoustic wave that oscillates in the $x^3$ direction. These modes
belong to the sector of vector cosmological perturbations.

There are three independent combinations of components of $h$
left, namely $h_{01}$, $h_{11}$ and $h_{kk}$, describing scalar
cosmological perturbations. But they are all inflicted by the
remaining freedom of choosing the zero time hypersurface. Pure
shift of this hypersurface, corresponding to a plane wave with
the wave covector $k = (k, 0, 0)$, yields a contribution to $h$
proportional to $h_{01} = 1$, $h_{11} = h_{22} = h_{33} =
2\dot{z}$. To identify physically relevant perturbations, we
define combinations of components of $h$ that are invariant
with respect to this hypersurface change. We do this by
introducing $y_{01}$, $y_{11}$ and $y$ such that
$$h_{01} =y_{01} +y,$$
$$h_{11} =y_{11} +2\dot{z}y,$$
$$h_{kk} =y_{11} +6\dot{z}y.$$

The interpretation of these formulas is as follows. If $\dot{z}
\neq 0$, we can always choose the zero time hypersurface in such a
way that $h_{kk} =h_{11}$. For such hypersurface, $y_{01}$ and
$y_{11}$ are equal to $h_{01}$ and $h_{11}$ respectively. If we
want to choose another hypersurface, we simply add the
hypersurface shifting terms proportional to $y$. Since we still
want $t$ to be the proper time of particles, we will have
differential equation for $y$, but we can see that the physical
information about the longitudal acoustic waves is contained in
$y_{01}$ and $y_{11}$ only.

Using the definition of $y$'s, we obtain a set of equations that
can be neaten to one equation for $y$,
\begin{equation}
\dot{y} = -\frac{1}{4\dot{z}}e^{-3z}(\ep +\si)y_{01}, \label{time}
\end{equation}
and a pair of equations not containing $y$ which invariantly
describe the propagation of a longitudal acoustic wave,
\begin{equation}
\dot{y}_{01} = \left[\frac{2\mu +3\la +5\si}{\ep +\si}\dot{z}
+\frac{1}{4\dot{z}}e^{-3z}(\ep +\si)\right]y_{01} -\po\frac{\la
+2\mu +3\si}{\ep +\si}y_{11}, \label{long1}
\end{equation}
\begin{equation}
\dot{y}_{11} =\left[2e^{-2z}k^2 +\frac{3}{2}e^{-3z}(\ep
+\si)\right]y_{01} -\frac{1}{4\dot{z}}e^{-3z}(\ep +\si)y_{11}.
\label{long2}
\end{equation}

\section{Example}\label{e}
Consider a continuum with the stiff ultrarigid state function
introduced in \cite{KS3}. The authors define the principal linear
particle densities $n_1$, $n_2$, $n_3$ in terms of $H^{AB}$, and
write the state function as
$$\rho = A\left[(n_1 n_2)^2 +(n_2 n_3)^2 +(n_3 n_1)^2\right] + B,$$
where $A$ and $B$ are constants. If we choose, for simplicity, $A
= 1$ and $B = 0$, the state function as a function of $H^{AB}$ is
\begin{equation}
\ro = \po(H^{CC}H^{DD} -H^{CD}H^{CD}). \label{stiff}
\end{equation}
If we treat all $H^{AB}$'s as truly independent variables, the
second term depends on the antisymmetric part of $H^{AB}$;
however, there is no need to modify it as long as we use our
formalism of symmetrized higher derivatives.

The first and second derivatives of $\ro$ with respect to $H^{AB}$
are
$$\Pa{\ro}{H^{AB}} = \de_{AB}H^{CC} -\de_{AC}\de_{BD}H^{CD},$$
$$\Ppa{\ro}{H^{AB}}{H^{CD}} = \de_{AB}\de_{CD} -\de_{AC}\de_{BD}.$$
Evaluating $\ro$ and its derivatives for $H^{AB} =e^{-2z}\de^{AB}$
we get
$$\ro = 3e^{-4z},\ \ \ \ \ \ \ \ \Pa{\ro}{H^{AB}} =
2e^{-2z}\de_{AB},\ \ \ \ \ \ \ \ \Ppa{\ro}{H^{AB}}{H^{CD}} =
\de_{AB}\de_{CD} -\de_{AC}\de_{BD},$$ and, consequently,
\begin{equation}
\rho_0 = 0,\ \ \ \ \ \ \ \ \ep = 3e^{-z},\ \ \ \ \ \ \ \ \si =
e^{-z},\ \ \ \ \ \ \ \ \la = -e^{-z},\ \ \ \ \ \ \ \ \mu = e^{-z}.
\label{ro}
\end{equation}
The equation $6\dot{z}^2 = \rho_0 +e^{-3z}\ep$ reads
$$\dot{z}^2 = \po e^{-4z}$$
and has an (expanding) solution
\begin{equation}
z = \frac{1}{4}\ln 2 +\po\ln t. \label{expansion}
\end{equation}
This is just the well known formula $a \propto \sqrt{t}$ for a
universe filled with an ideal fluid with the equation of state
$\sigma = (1/3) \ep$ (or, in more common notations, $p = (1/3)
\rho$).

Using $\dot{z} = (2t)^{-1}$ and $e^{-2z} = (\sqrt{2}t)^{-1}$, we
can express coefficients in the equation, or system of equations,
for each kind of waves as functions of time. The equation for
gravitational waves reads
\begin{equation}
\ddot h^T_{ij} +\frac{3}{2t}\dot h^T_{ij}
+\left(\frac{k^2}{\sqrt{2}t} +\frac{1}{t^2}\right)h^T_{ij} = 0.
\label{stgrav}
\end{equation}
Besides of that, we have a pair of equations for transversal
acoustic waves,
$$\dot{h}_{0\alpha} = \frac{1}{2t}h_{0\alpha} -\po h_{1\alpha},
\ \ \ \ \ \ \ \ \dot{h}_{1\alpha} = \left(\frac{k^2}{\sqrt{2}t}
+\frac{2}{t^2}\right)h_{0\alpha},$$ and a triplet of equations for
longitudal acoustic waves,
$$\dot{y} = -\frac{2}{t}y_{01},$$
$$\dot{y}_{01} = \frac{3}{2t}y_{01} -\po y_{11},\ \ \ \ \ \ \ \
\dot{y}_{11} =\left(\frac{\sqrt{2}}{t}k^2
+\frac{3}{t^2}\right)y_{01} -\frac{1}{t}y_{11}.$$ The first-order
equations for $h_{0\alpha}$ and $h_{1\alpha}$, as well as those
for $y_{01}$ and $y_{11}$, can be combined into a second-order
equation for either variable. We choose the variables
$h_{0\alpha}$ and $y_{01}$, because the corresponding equations
contain $k^2$ only as a factor in front of the variable itself,
thus becoming hyperbolic partial differential equations when one
passes from $k^2$ back to the Laplacian. The equations read
\begin{equation}
\ddot h_{0\alpha} - \frac{1}{2t}\dot h_{0\alpha}
+\left(\frac{k^2}{2\sqrt{2}t} +\frac{3}{2t^2}\right) h_{0\alpha} = 0,
\label{sttrans}
\end{equation}
\begin{equation}
\ddot y_{01} - \frac{1}{2t}\dot y_{01}
+\left(\frac{k^2}{\sqrt{2}t} +\frac{3}{2t^2}\right) y_{01} = 0.
\label{stlong}
\end{equation}
A remarkable feature of these equations is that they coincide in
the limit of small $k$. Thus, if we use for the description of
longitudinal and transversal acoustic waves the functions
$h_{0\alpha}$ and $y_{01}$, the waves behave identically in the
regime in which they stay well outside the horizon. There is no
point in extending the comparison to gravitational waves, since
the field $h^T_{ij}$ in the corresponding equation is of
completely different nature than the fields $h_{0\alpha}$ and
$y_{01}$ in the equations for acoustic waves (which are at least
akin). Note, however, that if we pick the variables $h_{1\alpha}$
and $y_{11}$ instead of $h_{0\alpha}$ and $y_{01}$ to describe
acoustic waves, or multiply $h_{0\alpha}$ and $y_{01}$ by
$e^{-2z}$ (the same factor that appears in the definition of
$h_{ij}$ in terms of $\delta g_{ij}$), the resulting equations
will coincide not only with each other, but also with the equation
for $h^T_{ij}$ in the limit of small $k$.

\begin{figure}[ht]
\centerline{\includegraphics[width=12cm]{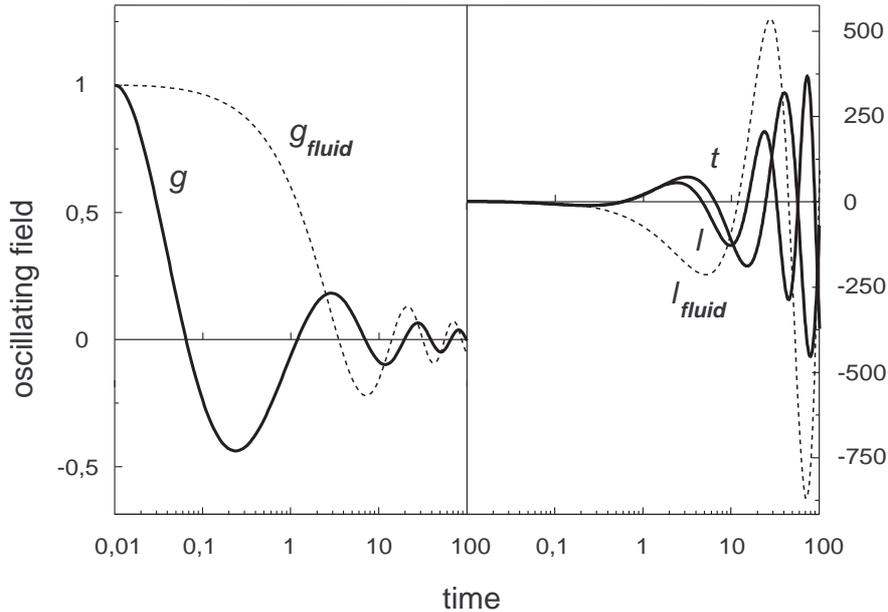}}
\caption{\small Time evolution of gravitational and acoustic
waves} \label{fig:relast}
\end{figure}

For both gravitational and acoustic waves, oscillating fields as
functions of time are depicted in figure \ref{fig:relast}. The
curves denoted by $g$ ('gravitational'), $t$ ('transversal') and
$l$ ('longitudinal') are graphs of the functions $h^T_{22}$ or
$h^T_{23}$, $h_{02}$ or $h_{03}$, and $y_{01}$ respectively. For
comparison, graphs of the functions $h^T_{22}$ or $h^T_{23}$ and
$y_{01}$ for an ideal fluid with the equation of state $\sigma =
(1/3) \ep$ are included into the figure, too (denoted by
$g_{fluid}$ and $l_{fluid}$; of course, there are no transversal
acoustic waves in an ideal fluid). All curves are computed for $k
= 1$, with the oscillating field equal to 1 and its derivative
vanishing at the moment $t = 0.01$.
The waves cross the horizon at $t \sim 1$; before that their
wavelength is larger and afterwards it is smaller than the horizon
length. The main new effect of nonzero shear stresses,
distinguishing the continuum under consideration from an ideal
fluid, are oscillations of the fields at superhorizon scales.
Asymptotically, the oscillating factor is $\cos[(\sqrt{15}/4) \log
t + \psi]$ for all three kinds of waves.

\section{Conclusion}\label{z}
We have developed a formalism describing the propagation of plane
waves in a general relativistic homogeneous and isotropic flat
continuum. In comparison with \cite {SDM}, where the same problem
is studied, we did not include into the theory small perturbations
to the Euclidean metric of the material manifold (tensor $b_{ab}$
in the notation of \cite {SDM}). In this way we simplified the
matters as much as possible.

We were successful in identifying relativistic generalizations of
acoustic waves. The longitudinal mode was encumbered with the
'time wave' gauge freedom, and the definition of gauge invariant
$y$'s was needed. After introducing them we were able to describe
the wave by two coupled first-order differential equations. This
is equivalent to one second-order equation, hence our theory is a
straightforward generalization of its nonrelativistic counterpart.
Again, this can be compared with \cite {SDM} where the propagation
of the longitudinal wave is described by two coupled {\it
second-order} equations (equations (3.12) and (3.15) in subsection
3.1). The cited work differs from ours in that respect that it
uses Newtonian gauge. Of course, the transition from one gauge to
another cannot increase the number of degrees of freedom,
therefore in Newtonian gauge the longitudinal wave must be also
described by two first-order equations (imposing constraints on
initial conditions for the second-order equations). Perhaps the
comoving proper-time gauge we have chosen for our calculations is
better adapted for derivation of these more fundamental equations.

We kept the formalism as easy and simple as possible, although
sometimes we have preferred comprehensibility to simplicity. For
example, the introduction of $\rho_0$ and $\ep$ was not necessary.
We could use $\ro$ (and scalars entering the derivatives of $\ro$
with respect to $H^{AB}$) only, but we felt that the quantities
$\ep$ (and $\si$, $\mu$ and $\la$) were more appropriate for the
description of the continuum. Also, we introduced $z$ and $h$'s
just to simplify the expressions even if the meaning of $h_{0i}$
is far from obvious. And our choice of $y$'s reflected the most
natural initial condition $h_{11} = h_{kk}$. On the other hand,
the fact that the resulting equations do have singularity for
$\dot{z} = 0$ suggests that this condition is not the best. For
those interested in nonsingular equations, note that using
$$h_{01}= \yp_{01} +\yp,$$
$$h_{kk}= \yp_{kk} +6\dot{z}\yp,$$
$$h_{11}= \frac{4e^{-2z}k^2 +2e^{-3z}(\ep +\si)}{4e^{-2z}k^2}\yp_{kk}
+2\dot{z}\yp,$$ we obtain equations
$$\dot{\yp}= -\frac{e^{-3z}(\ep +2\mu +3\la +6\si)}{4e^{-2z}k^2
+3e^{-3z}(\ep +\si)}\yp_{kk},$$
$$\dot{\yp}_{kk}= \frac{6e^{-3z}(\ep +2\mu +3\la
+6\si)}{4e^{-2z}k^2 +3e^{-3z}(\ep +\si)}\dot{z}\yp_{kk}
+2e^{-2z}k^2\yp_{01},$$
$$\dot{\yp}_{01} = \frac{2\mu +3\la +5\si}{\ep +\si}\dot{z}\yp_{01}
-\left[\po\frac{2\mu +\la +3\si}{\ep +\si} +\frac{2e^{-3z}(\mu
+\si)}{4e^{-2z}k^2} -\frac{e^{-3z}(\ep +2\mu +3\la
+6\si)}{4e^{-2z}k^2 +3e^{-3z}(\ep +\si)}\right]\yp_{kk}.$$
However, these equations look no simpler than the equations for
$h$'s, and we used $\dot{z} \neq 0$ when deriving them anyway.

When applying our theory to a stiff ultrarigid continuum, we found
that the time evolution of longitudinal and transversal acoustic
waves, if described by the functions $h_{0\alpha}$ and $y_{01}$,
coincide in the limit of small $k$. This is not a unique feature
of the state function considered. The same statement holds for any
'linear' continuum, that is, continuum with $\rho_0 = 0$ and
$\si$, $\la$ and $\mu$ proportional to $\ep$, provided the
constraint on the coefficients of proportionality following from
the relation ${d\si}/{dz} = -(2\si +3\la +2\mu)$ is satisfied.

{\it Acknowledgement.} This work was supported by the grant VEGA
1/3042/06.

\end{document}